\documentclass[a4paper,11pt]{article}
\usepackage{pos}

\title{Non-resonant Streaming Instability driven by Leptons}
 \ShortTitle{Lepton-driven Bell instability}

\author*[a]{Siddhartha Gupta}
\author[a,b]{Damiano Caprioli}
\author[c]{Colby Haggerty}

\affiliation[a]{Department of Astronomy and Astrophysics, The University of Chicago, 5640 S Ellis Ave, Chicago, IL 60637, USA}
\affiliation[b]{Enrico Fermi Institute, The University of Chicago, 5640 S Ellis Ave, Chicago, IL 60637, USA}
\affiliation[c]{University of Hawaii, Honolulu, HI, 96822, USA}

\emailAdd{gsiddhartha@uchicago.edu}

\abstract{
Using fully-kinetic plasma simulations, we study the non-resonant (Bell) streaming instability driven by energetic leptons. We identify the necessary conditions to drive it and the differences from the standard proton-driven case in both linear and saturated stages. 
A simple analytic theory is presented to explain simulations. Our findings are crucial for understanding the phenomenology of astrophysical environments where only electrons may be accelerated (e.g., oblique shocks) or where relativistic pairs are produced (e.g., around pulsar wind nebulae).}

\FullConference{37$^{\rm{th}}$ International Cosmic Ray Conference (ICRC 2021)\\
		July 12th -- 23rd, 2021\\
		Online -- Berlin, Germany}

\begin{document}
\maketitle

%
%
%
\section{Introduction}
%
%
%
In many circumstances, charged particles drifting in the plasma with super-Alfv\'{e}nic speed trigger plasma instabilities that create electromagnetic fluctuations. 
One such mechanism produces modes that are non-resonant with the driving particles and leads to $ \delta B/ B_{\rm 0}\gg 1$, where $B_{\rm 0}$ is the initial magnetic field. 
This non-resonant streaming instability (hereafter NRSI) was proposed in \citep{lucek+00,bell04} to explain the observations of synchrotron X-ray emissions in astrophysics shocks \citep[e.g.,][]{parizot+06}. 
Later several studies have investigated the NRSI per se \citep[e.g.,][]{niemiec+08,ohira+09,riquelme+09,amato+09}, or in the context of cosmic ray (CR) acceleration at shocks \citep[e.g.,][]{caprioli+14b} and propagation in galactic and extra-galactic environments \citep[e.g.,][]{blasi+15,schroer+21}.
Finally, there is an ongoing effort to study NRSI in laboratory \citep[e.g.,][]{jao+19}.
 
Kinetic studies of the NRSI have included the effects of the background plasma temperature \citep[e.g.,][]{reville+08a,zweibel+10,marret+21} and the shape of the CR distribution function \citep[e.g.,][]{haggerty+19p}, always considering a current of energetic protons. 
Although protons are more abundant than electrons in CRs, there are environments where leptonic currents can be expected; they include:
1) quasi-perpendicular shocks, where reflected electrons can support a current in the shock upstream \citep{guo+14a,xu+20}, which may apply, e.g., to the  termination shocks from the winds of stellar clusters \citep[][]{gupta+18};
2) around pulsar wind nebulae (PWNe),  where both electrons and positrons are accelerated \citep[e.g.,][references therein]{philippov+18,hawc17};
3) the electron strahl population in the solar wind \citep[e.g.,][]{halekas+19}. 

Here we study the lepton driven NRSI and compare its characteristics with the ion-driven case, including the linear growth, the spectrum, the structure and the saturation of the magnetic field. 
We used a semi-classical analytic theory to examine the lepton driven NRSI and verified these predictions with kinetic simulations, presented in \citep{gupta+21}.

The wavenumber and the growth rate of the fastest growing mode read \citep{bell04,amato+09}:
\begin{equation}
k_{\rm fast} = \frac{1}{2}\frac{n_{\rm cr}}{n}\frac{v_{\rm d}}{v_{\rm A0}} \frac{1}{d_{\rm i}}\,\ \ {\rm and} \,\  \gamma_{\rm fast} = \frac{1}{2}\frac{n_{\rm cr}}{n}\frac{v_{\rm d}}{v_{\rm A0}} \omega_{\rm ci} \, , 
\end{equation}
where $v_{\rm A0}$ is the Alfv\'{e}n speed and $v_{\rm d}$ is the drift velocity of the driving particles in the beam (hereafter, CRs); 
$n_{\rm cr}$ and $n$ are the particle density in the CR beam and background plasma. 
For typical astrophysical shock parameters, the wavelength and corresponding growth timescale are
\begin{equation}
\lambda_{\rm fast} \approx 10^{-4}{\rm pc}\, \left(\frac{n_{\rm cr}/n}{10^{-7}}\right)^{-1}\left(\frac{v_{\rm d}/v_{\rm A0}}{10}\right)^{-1}\left(\frac{n}{\rm cm^{-3}}\right)^{-1/2}\, {\rm and}\ \tau_{\rm fast} \approx 1\, {\rm yr} \,\left(\frac{n_{\rm cr}/n}{10^{-7}}\right)^{-1}\left(\frac{v_{\rm d}}{10^7{\rm cm\,s^{-1}}}\right)^{-1}\left(\frac{n}{\rm cm^{-3}}\right)^{-1/2}
\end{equation}
which suggest that the NRSI can be relevant in many environments and that it often needs to be treated with sub-grid prescriptions that account for small length/time scales.  

The main conditions to drive the instability are found as
\begin{eqnarray}
v_{\rm d}& \gg  & v_{\rm A0}\\
\gamma_{\rm fast}& \ll  & \omega_{\rm ci}\\
\xi\equiv \frac{n_{\rm cr}}{n} \frac{p_{\rm cr}}{m_{\rm i}}\frac{v_{\rm d}}{v_{\rm A0}^2} \equiv \frac{1}{2}\frac{P_{\rm cr}}{P_{\rm B0}} & \gg  & 1 \ .
\end{eqnarray}
The first condition demands the super-Alfv\'{e}nic drift of CRs in the plasma frame, analog to the resonant streaming instability \citep[][]{kulsrud+68}. 
The second one imposes the growth rate to be smaller than the ion cyclotron frequency (magnetization condition), which can be interpreted as the current density in the driving beam, $J_{\rm cr}\propto \,n_{\rm cr} v_{\rm d}$, to be smaller than the equivalent Alfv\'{e}nic current, $J_{\rm A0}\propto n\, v_{\rm A0}$ in the background plasma. 
The third condition is equivalent to $\lambda_{\rm fast} \ll R_{\rm L,cr}$, i.e., the most unstable mode are much smaller than the CR gyroradius;
this is achieved when the net CR momentum flux, $P_{\rm cr}$, is much larger than the magnetic pressure, $P_{\rm B0}$. 
For an arbitrary charge/mass of particles in the CR beam, the dependence of $\xi$ on $p_{\rm cr}/m_{\rm i}$ suggests that for lepton driven NRSI, leptons must have large Lorentz factor to get a similar result of ion-driven case.
For leptons, in principle one has also to require the growth rate of the instability to be larger than the synchrotron loss rate \citep[see \S2 of][]{gupta+21}. Below we briefly discuss our main results.
%
%
%
\section{Simulation setup and Results}
%
%
%
We have used a particle-in-cell code, {\bf Tristan-MP} \citep{spitkovsky05} to simulate the NRSI.
We use quasi-1D boxes, but retain all three components of vectors. The number of cell along the $x$-direction is set in way that the domain length $\gtrsim 6 \lambda_{\rm fast}$. For all runs, the grid spacing $\Delta_{\rm x} = 0.2\,d_{\rm e}$ and the time step $\Delta_{\rm t} = 0.04\, \omega_{\rm pe}^{-1}$. 
The magnetization is chosen such that the Alfv\'{e}n speed covers a range from $3.2\times 10^{-3} c $ to $4\times 10^{-2} c$ for different runs. 
The velocity distribution of background plasma is set to Maxwellian where $T_{\rm i}= T_{\rm e}$ and the ion thermal speed $a_{\rm i0}\equiv(k_{\rm B} T_{\rm i}/m_{\rm i})^{1/2}\approx 8\times 10^{-3} c$, i.e., the initial plasma-$\beta \equiv P_{\rm g0}/P_{\rm B0} \sim 1$. 
The CR distribution is isotropic in its rest frame, where particles have a Lorentz factor $\gamma_{\rm cr}$. The number of particles per cell  is set to $25$ per species and their weights are modified to ensure charge neutrality and to set the different values of $n_{\rm cr}/n$. For more details, see section 3 in \cite{gupta+21}.
%
\subsection{Characteristics of the growing B-fields}
\begin{figure}[ht!]
\begin{minipage}{0.5\textwidth}
\centering
\includegraphics[width=1.\linewidth, height=0.95\linewidth]{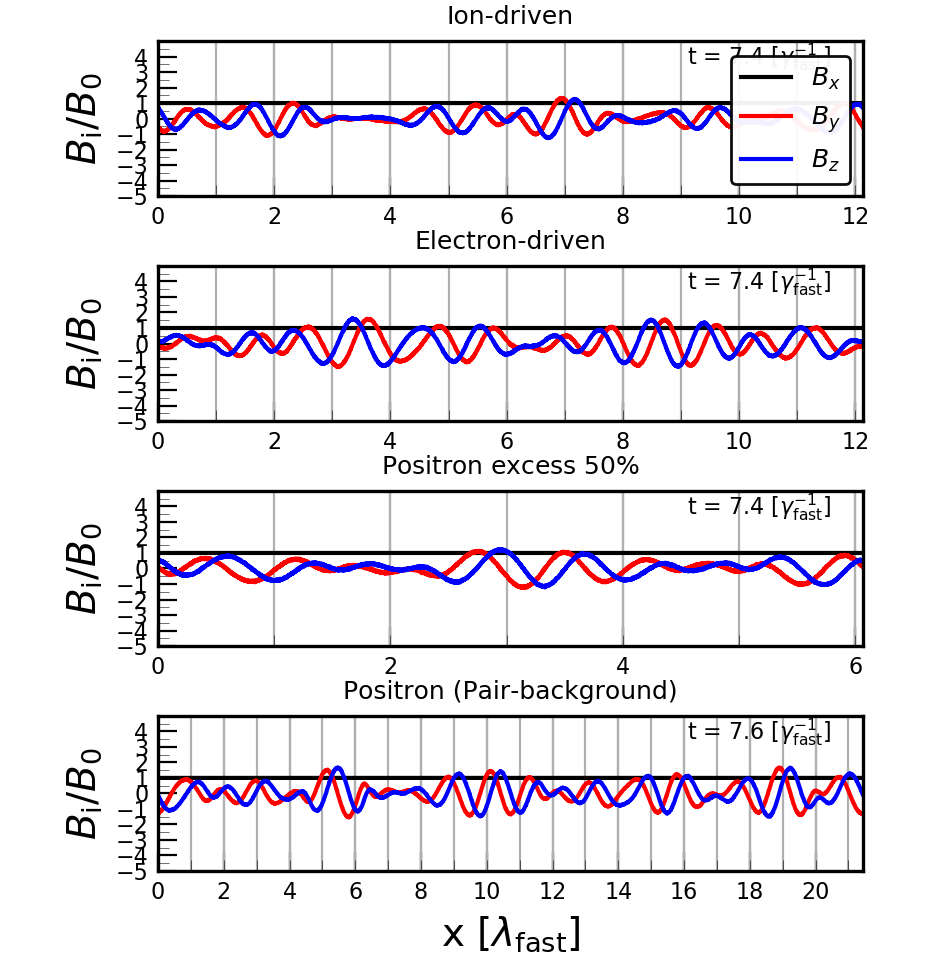}
\end{minipage}
\begin{minipage}{0.5\textwidth}
\centering
\includegraphics[width=0.9\linewidth, height=0.95\linewidth]{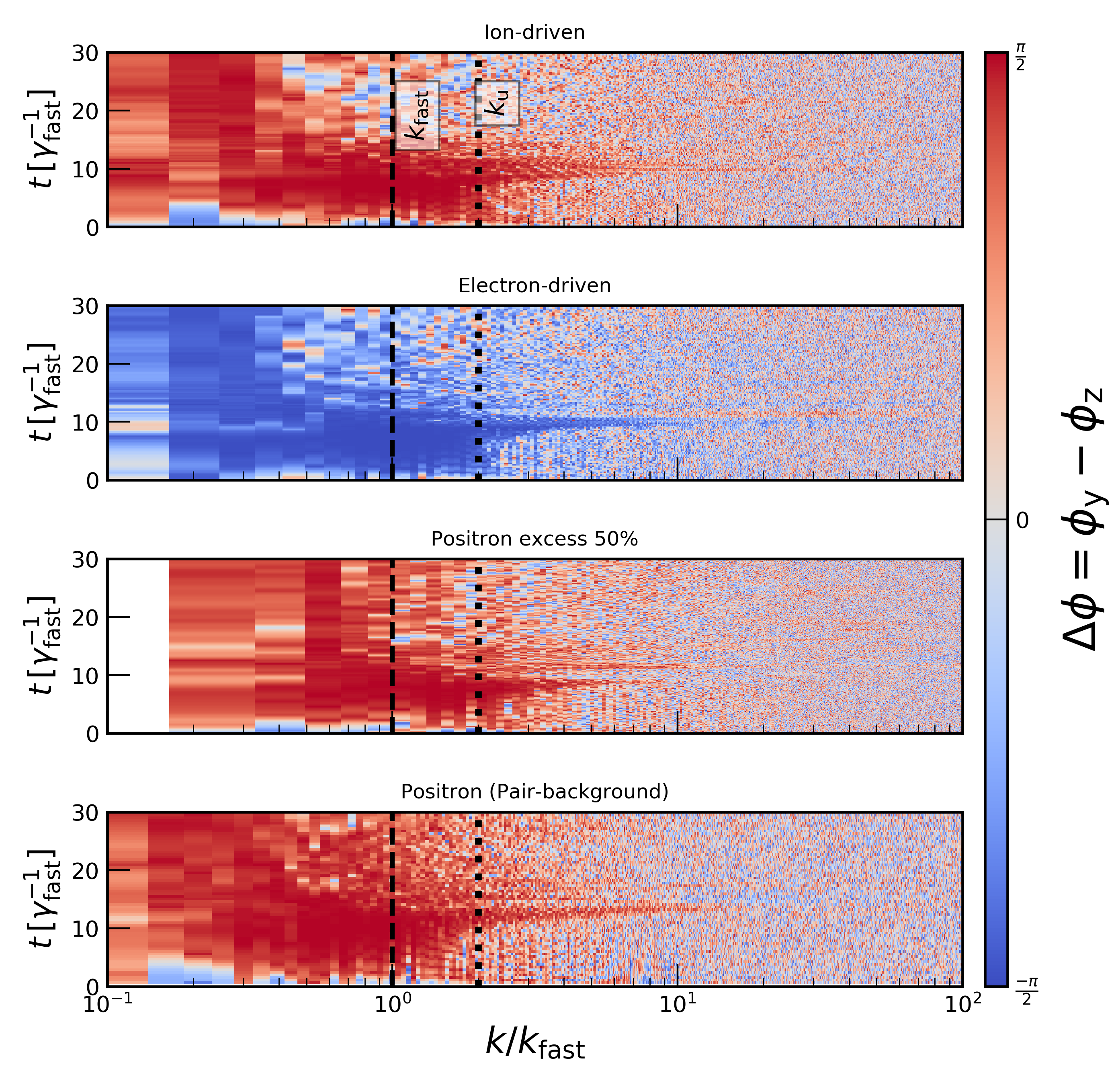}
\end{minipage}
\caption{
Left panels: Simulated profiles of $B_{\rm x}, B_{\rm y},$ and $B_{\rm z}$ at $t\approx 7.5\,\gamma_{\rm fast}^{-1}$, representing the development of the NRSI in four different environments;
the wavelength of the fastest growing mode follows the analytic expectation. 
Right panels: the phase difference between $B_{\rm y}$ and $B_{\rm z}$ at different modes for all four cases. From this figure it is evident that the negatively-charged beam drives left-handed ($\Delta \phi <0$ ) modes.}
\label{fig:profiles}
\end{figure}

%
The left panels of Figure \ref{fig:profiles} show the profiles of magnetic fields from four selected simulations. 
All panels except the third show cases where the CR beams contain a single species, i.e., either ion or electrons or positron. 
The second panel shows the electron-driven case. 
In the third panel, the CR beam is made of pair plasma but positron number density is larger than that of electrons, making the beam positively charged with $50\%$ positron excess. 
$10\%$ and $20\%$ positron excess cases lead to similar results \citep[][]{gupta+21}. 
In all panels, the wavelength of the dominant modes is consistent with the linear theory, i.e., it depends on the effective current density in the CR beam.
 
To better characterize the unstable modes, we analyze the phase difference between $B_{\rm y}$ and $B_{\rm z}$ by taking the Fourier transform of $B_{\rm y}$ and $B_{\rm z}$ along the $x$-axis and introducing 
\begin{equation}
   \Delta\phi (k) \equiv \phi_{\rm y, k}-\phi_{\rm z, k} = -\tan^{-1}\left(\frac{V_{\rm k}}{\sqrt{U^2_{\rm k}+Q^2_{\rm k}}}\right)\, .
\end{equation}
Here $Q_{\rm k}, U_{\rm k}$ and $V_{\rm k}$ are the Stokes parameters at a given mode $k$.
The right panels of Figure \ref{fig:profiles} show the time evolution of $\Delta\phi$ as a function of $k$ for the same runs discussed above. 
Note that for $k< 2 k_{\rm fast}$, the value of $\Delta\phi (k)$ is definite:
growing modes are left-handed (right-handed) when the CR beam is negatively(positively)-charged. 
After $t\approx 10\,\gamma_{\rm fast}^{-1}$ non-linear effects cannot be ignored. 
%
\subsection{Back-reaction in the Non-linear Stage} \label{subsec:bck}
%
The departure from the linear theory is due to the response of both the background plasma and the CR beam to the growing electromagnetic fields. 
\begin{figure}[hb!]
\centering
\includegraphics[scale=0.27]{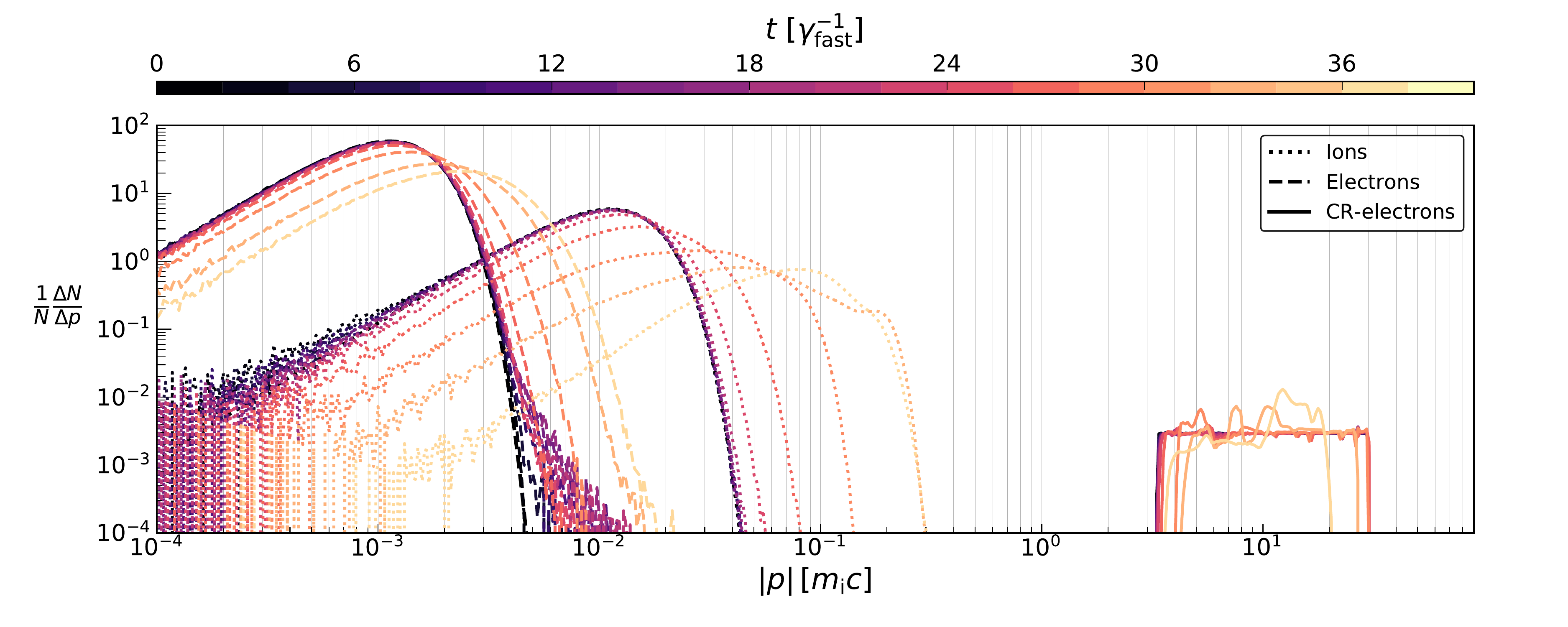}
\caption{Spectra of CR electrons (solid curves), and thermal electrons (dashed curves) and ions (dotted curves) at different times in the simulation frame. Figure shows CRs lose their momentum with time, while thermal background gains, as expected in the momentum conservation.}\label{fig:distribution}
\end{figure}
This back-reaction can be observed in Figure \ref{fig:distribution}, which displays how the distribution functions evolve with time. 
Note the momentum/energy of driving particles (in this case energetic electrons) transfers to the background plasma. 
For all lepton-driven NRSI cases, we have seen a similar evolution of distribution function of different species.
We have performed some diagnostics to quantify such modifications, as outlined below.

\begin{figure*}
\begin{minipage}{0.48\textwidth}
\centering
\includegraphics[width=0.9\linewidth, height=0.8\linewidth]{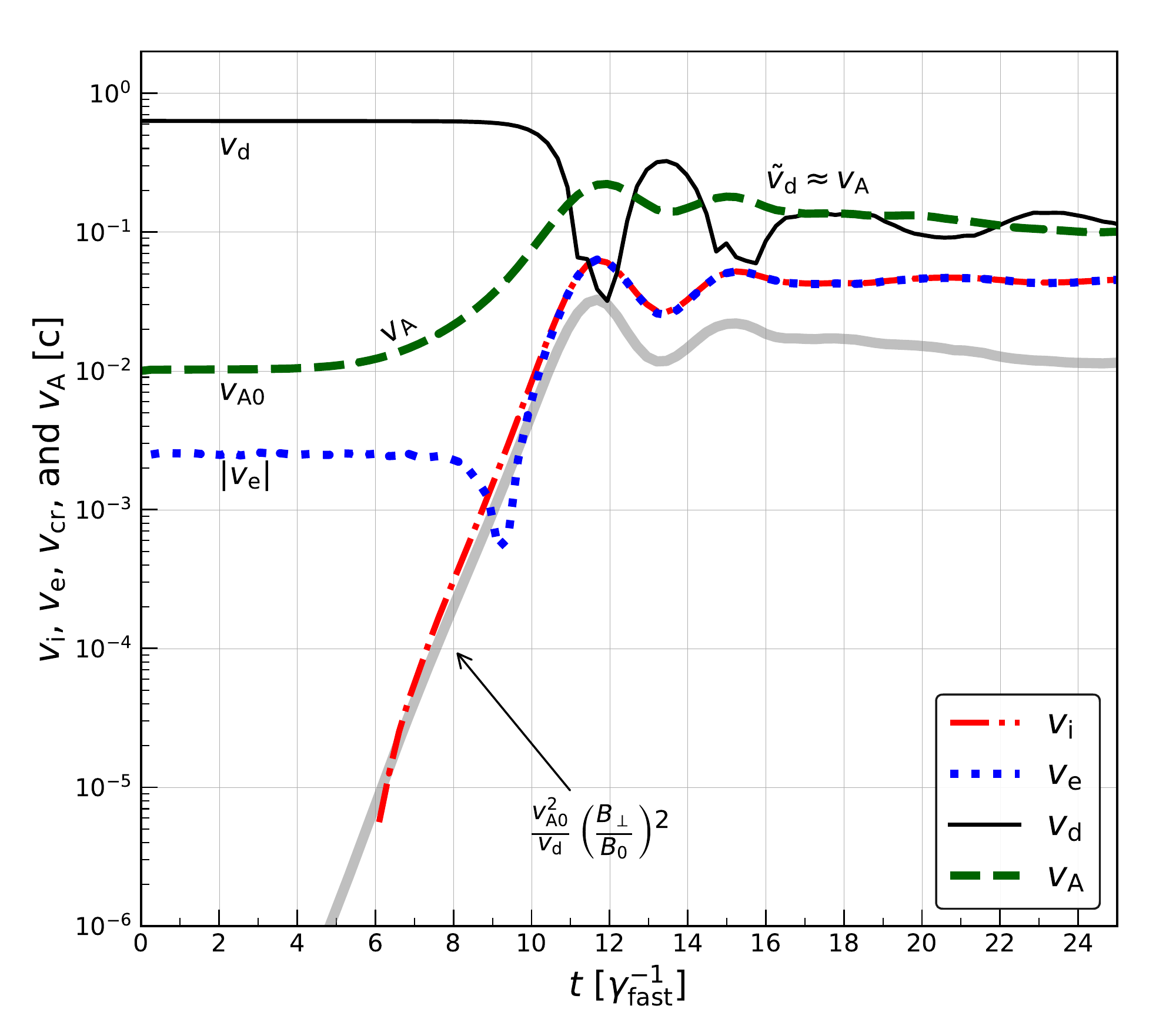}
\end{minipage}
\begin{minipage}{0.48\textwidth}
\centering
\includegraphics[width=0.9\linewidth, height=0.8\linewidth]{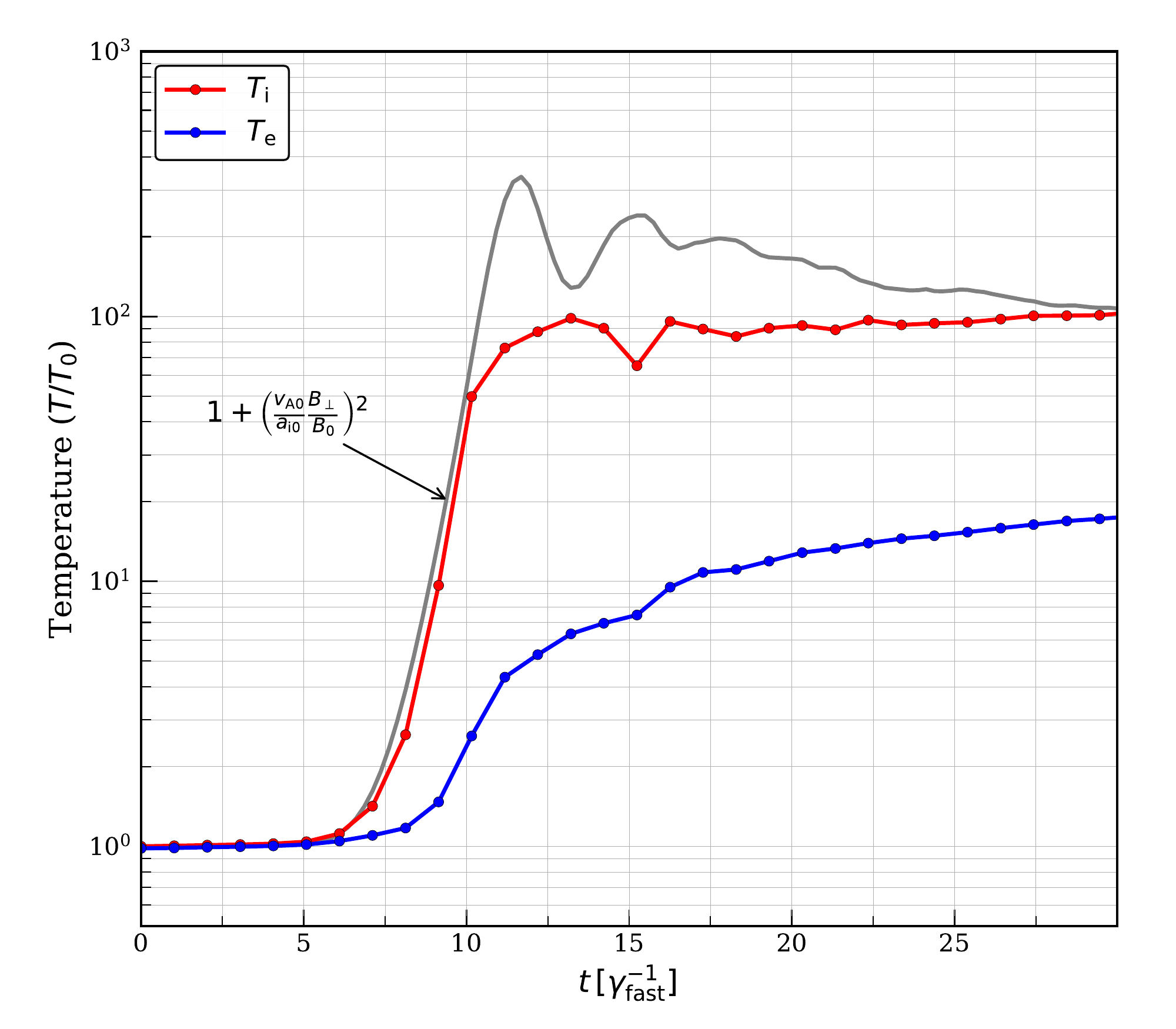}
\end{minipage}
\caption{Left panel: box averaged drift velocity along the $x$ axis for different species, and the Alfv\'{e}n speed. At final stage the drift velocity of CRs modifies to the amplified Alfv\'{e}n speed. Right panel: electron and ion temperature (calculated using Equation \ref{eq:tmp}) at different times. The grey curves in both panels represent analytic expectations (detailed in Appendix A.2 in \cite{gupta+21}).}\label{fig:compvt}
\end{figure*}

At first we check how the drift velocity of different species changes with time. 
We consider the average of the $x$-component of the velocity, which gives the bulk speed of a species in the simulation frame.
In Figure \ref{fig:compvt} (left panel), the black and blue/red curves denote the drift velocity of CRs and thermal electrons/ions (represents the run EI-S-$\xi$340 with CR electrons, detailed in Table 1 of \cite{gupta+21}).
Until $t\approx 9\,\gamma_{\rm fast}^{-1}$, $v_{\rm d}\equiv v_{\rm crx}=0.635 c$ and $|v_{\rm ex}|\simeq 0.0025$ remain close to their initial values (\S3 in \cite{gupta+21}). 
The evolution of red and blue curves is consistent with the linear prediction, shown by the grey curve. 
As the B-field increases, the Alfv\'{e}n speed also increases and we find that when $t>10\,\gamma_{\rm fast}^{-1}$, $v_{\rm crx}\approx v_{\rm A}$. 
This implies that in the final stage of NRSI, CRs drift with the amplified Alfv\'{e}n speed. 
This is an important feature, which can be used to predict the final $B_{\rm \perp}$ at saturation \citep[e.g., Appendix A.2 in][]{gupta+21}.

We also estimate the effective temperature ($T$) of ions and electrons at different times by taking the second moment of their distributions,
\begin{equation} \label{eq:tmp}
T_{\rm \alpha,rs} \equiv \frac{1}{N_{\rm \alpha}}\sum_{p=1}^{N_{\rm \alpha}} m_{\rm \alpha}\,(u_{\rm \alpha r}^{\rm p}-v_{\rm \alpha r})(u_{\rm  \alpha s}^{\rm p}-v_{\rm  \alpha s})\,, \  \text{where $r,s\in x,y,z$}
\end{equation}
Here $N_{\rm \alpha}$ represents the total number of macro-particles for each species ($\alpha= i/e$) in computational domain, and $u_{\rm \alpha r}$ is the velocity component of a particle along the direction $r$.
In right panel of Figure \ref{fig:compvt}, the red and blue curves denote ion and electron temperatures (normalized to the initial temperature $T_{\rm i}=T_{\rm e}$);
the temperature of the plasma rapidly increases during $t\approx 7-15\,\gamma_{\rm fast}^{-1}$, along with the B-field (left panel). 
The increase in the kinetic energy and temperature of the plasma can be understood from Equations A16 and A17 in \cite{gupta+21}, which give the final temperature of the plasma:
\begin{equation}\label{eq:tmp}
  T_{\rm \alpha}\sim T_{\rm \alpha0} \left[1+\left(\frac{v_{\rm A0}}{a_{\rm \alpha0}}\right)^2\left(\frac{B_{\rm \perp}}{B_{\rm 0}}\right)^2\right]
\end{equation}
For our fiducial run, $v_{\rm A0}=10^{-2}\,c$, $a_{\rm i0}\approx 8\times 10^{-3}\,c$, and at the saturation $B_{\rm \perp}/B_{\rm 0}\approx 10$ (cf. Figure \ref{fig:sat} the run EI-S-$\xi$340), Equation \ref{eq:tmp} gives $T_{\rm i}/T_{\rm i0}\sim 150$ and this qualitatively agrees with the red curve in the right panel of Figure \ref{fig:compvt} (representing temperature of the ions). 
The interpretation of the raise in electron temperature is more complicated than the ions because the non-linear damping of different modes may contribute significantly, which are not taken into account in our analytic calculation. 
%
\subsection{Saturation}
\begin{figure*}
\centering
\includegraphics[scale=0.45]{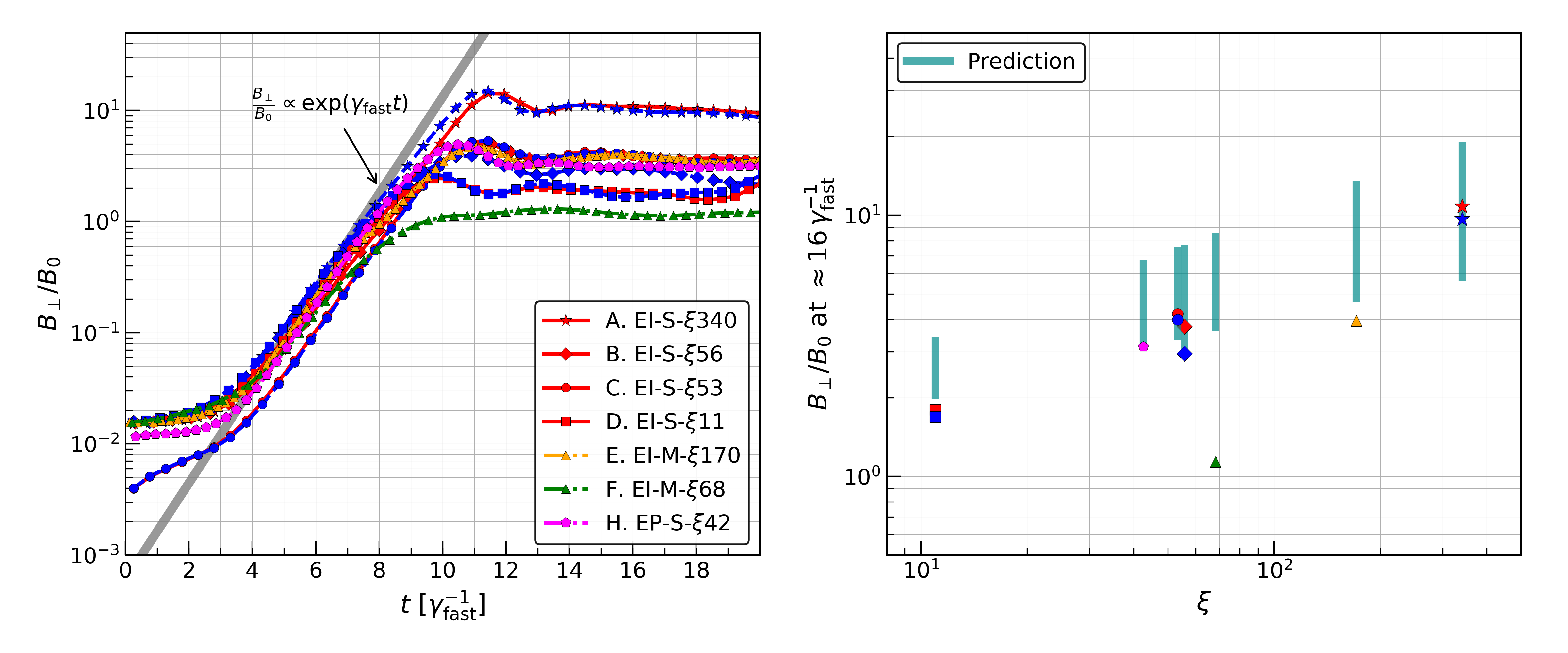}
\caption{ Time evolution of $B_{\rm \perp}=\sqrt{B_{\rm y}^2+B_{\rm z}^2}$ for different CR compositions, displaying all follow the linear prediction, marked by grey solid line. 
Right panel show the comparison between simulation (points) and analytic predictions (cyan lines). The parameters used in these simulations are detailed in Table 1 of \cite{gupta+21}.} \label{fig:sat}
\end{figure*}

%
We have also investigated the saturated value of the magnetic field. 
The box-averaged $B_{\rm \perp}=\sqrt{B_{\rm y}^2+B_{\rm z}^2}$ for all simulations with non-zero beam current are displayed in Figure \ref{fig:sat}. 
The left panel shows the linear growth of $B$ until $\approx 10\, \gamma_{\rm fast}^{-1}$, while the right panel in the same figure shows $B_{\rm \perp}/B_{\rm 0}$ at saturation. 
The cyan bars represent the range of $B_{\rm \perp}/B_{\rm 0}$ at saturation, obtained from semi-classical theory (see Appendix A.2 in \cite{gupta+21}):
\begin{eqnarray}
\frac{B_{\rm \perp}}{B_{\rm 0}} \approx
\begin{cases}
  \xi^{1/2} & \text{without accounting the heating losses} \\
  \left[\left(b^2+4\,\xi\right)^{1/2}-b\right]^{1/2} &  \text{with the heating loss corrections} 
\end{cases}
\end{eqnarray}
Here $b=2+\beta/2$ and $\beta\equiv P_{\rm g0}/P_{\rm B0}$ is the plasma beta. 
When the beam current is zero (i.e., $\xi \equiv 0$), both solutions yield $B_{\rm \perp}/B_{\rm \perp} \rightarrow 0$, as expected. 
The solution $B_{\rm \perp}/B_{\rm 0} \approx \xi^{1/2}$ corresponds to fields saturating when the magnetic pressure becomes comparable with the anisotropic CR pressure \citep{bell04,blasi+15} (see the contribution by \cite{zacharegkas+21p}). 
The other solution is important in hot plasmas, where the NRSI growth rate may be smaller \citep[see e.g.,][]{reville+08a,zweibel+10,marret+21} and the saturated B-field may be significantly reduced. 
For our choice of parameters $\beta\approx 1$ and the growth rate is unaffected but the final $B_{\rm \perp}/B_{\rm 0}$ slightly reduces due to the raise in the background plasma temperature (see the right panel of Figure \ref{fig:compvt}).

%
%
%
\section{Conclusions}
%
%
%
In  \cite{gupta+21} we used kinetic simulations to study the characteristics of the NRSI for different compositions of driving particles and background plasma.
We find that, when the momentum flux of the driving beam is much larger than the magnetic pressure, the NRSI grows faster than the resonant instability, regardless of the CR charge.
The helicity of the growing modes is determined by the effective CR charge; 
in particular, a positively(negatively)-charged beam drives right(left)-handed modes. 
The saturation of the NRSI mainly depends on the beam momentum flux, and neither on its current, nor on the charge/mass of the driving particles.

When the beam encompasses both species, the growth rate is defined by the net current and magnetic fields saturate below equipartition (e.g., green triangle in Figure \ref{fig:sat}).
Yet, non-negligible magnetic fluctuations are found also for beams with null current, which may be sufficient, e.g., for the increased scattering rate of CR leptons inferred around PWNe \citep[][]{hawc17}.

\bibliographystyle{JHEP}
\bibliography{Total}

%
%
%

\end{document}